On multi-soliton solutions of the Sine-Gordon equation in more than one space dimension

Yair Zarmi

Jacob Blaustein institutes for Desert Research

Ben-Gurion University of the Negev

Midreshet Ben-Gurion, 8499000 Israel

Abstract

The (1+1)-dimensional Sine-Gordon equation passes integrability tests commonly applied to nonlinear evolution equations. Its soliton solutions are obtained by a Hirota algorithm. In higher space-dimensions, the equation does not pass these tests.

In this paper, using no more than the relativistic kinematics of the tachyonic momentum vectors, from which the soliton solutions are constructed through the Hirota algorithm, the existence and classification of $N$-soliton solutions of the (1+2)- and (1+3)-dimensional equations for all $N \geq 1$ are presented.

In (1+2) dimensions, each multi-soliton solution propagates rigidly at one velocity. The solutions are divided into two subsets: Solutions whose velocities are lower than the speed of light ($c = 1$), or are greater than or equal to $c$.

In (1+3)-dimensions, multi-soliton solutions are characterized by spatial structure and velocity composition. The spatial structure is either planar (rotated (1+2)-dimensional solutions), or genuinely three-dimensional – branes. Some solutions, planar or branes, propagate rigidly at one velocity, which is lower than, equal to, or higher than $c$. A subset of the branes contains "hybrids", in which different clusters of solitons propagate at different velocities. Some velocities may be lower than $c$ but some must be equal to, or greater than $c$.

Finally, the speed of light cannot be approached from within the subset of slower-than-light solutions in both (1+2) and (1+3) dimensions.

# 1. Introduction
## 1.1 Open problems concerning the Sine-Gordon equation

The Sine-Gordon equation has attracted wide attention over the years in the description of classical and quantum mechanical phenomena [1-9], and within the framework of quantum-field theory [9-17]. Hirota had produced an algorithm for the construction of the soliton solutions of the (1+1) dimensional equation [18]. It was then shown that the equation is integrable [19].

In contradistinction, for decades, it has been known that, in (1+2) dimensions, the equation does not pass integrability tests that are traditionally applied to nonlinear evolution equations. It is not integrable within the framework of the Inverse-Scattering formalism [20] and it does not have the properties required for integrability in a Painlevé analysis [21-24]. Furthermore, in (1+2) dimensions, single- and two-soliton solutions could be constructed through the Hirota algorithm, but the attempt to construct a three-soliton solution encountered an obstacle [25]: For a three-soliton solution to exist, the parameter sets (three parameters for each soliton) from which the solution is constructed had to obey a constraint. A different constraint was found in the construction of multi-soliton solutions of the (1+3) dimensional Sine-Gordon equation [26]. Over the years, quite a few works (see, e.g., Refs. [27-32]) have approached the issue of the construction of soliton solitons in (1+2) or (1+3) dimensions. Still, the full algorithm is presented only in Refs. [25, 26].

However, none of Refs. [18, 25-32] exploits very useful properties of the parameters that enter the construction of multi-soliton solutions in (1+$n$) dimensions. Each soliton is associated with a set of (1+$n$) parameters. These parameters may be viewed as the components of a tachyonic momentum vector (a vector with $(\text{mass})^2 = -1$) in the (1+$n$)-dimensional Minkowski space.

In this paper, the kinematics of such vectors, in particular, their properties under Lorentz transformations in Minkowski space, are exploited. This allows for a demonstration of a physical na-ture of the existence of N-soliton solutions for any $N \geq 1$ in (1+2) and (1+3) dimensions of SG$n$, the Sine-Gordon equation in (1+$n$)-dimensions,

$$\partial_\mu \partial^\mu u + \sin u = 0, \quad \mu = 0,1,..,n, \quad n = 1,2,3 \ . \tag{1}$$

More important, this approach reveals the richness of possible solutions of SG$n$.

Section 1.2 presents a review of properties of the Hirota [18] soliton solutions of SG1. Section 1.3 summarizes the results presented in the paper. All the arguments presented here are based on the properties under Lorentz transformations of the tachyonic momentum vectors, from which soliton solutions are constructed through the Hirota Algorithm. Section 2 is devoted to a review of these properties. As simple as the considerations are, they have not been discussed in the literature, probably, owing to the lack of interest in the kinematics of space-like vectors. Sections 3 and 4 discuss the construction of soliton solutions in, respectively, (1+2) and (1+3) dimensions. Section 5 presents an invariance property of the subset of multi-soliton solutions of SG3, which propagate at a velocity that is lower than $c$. Section 6 briefly summarizes the results for Eq. (1), with the sign of the $\sin u$ -term changed to a (−) sign.

**1.2 Soliton solutions of SG1 - original Sine-Gordon equation in (1+1) dimensions**
**1.2.1 Construction of solutions**
The first step in the Hirota algorithm [18] for the construction of the soliton solutions of SG1 is a transformation, originally proposed in the cases of one- and two-soliton solutions [7,8]:

$$u(x;P) = 4\tan^{-1}\left[g(x;P)/f(x;P)\right] \ . \tag{2}$$

In Eq. (2),
$$P \equiv \left\{p^{(1)}, p^{(2)}, ..., p^{(N)}\right\} \ . \tag{3}$$

$x$ and $p^{(i)}$ are coordinate and momentum vectors in (1 + 1) dimensions.

The functions $g(x;P)$ and $f(x;P)$ are given by:

$$g(x;P) = \sum_{\substack{1 \leq n \leq N \\ n \text{ odd}}} \left( \sum_{1 \leq i_1 < \cdots < i_n \leq N} \left\{ \prod_{j=1}^{n} \varphi\left(x; p^{(i_j)}\right) \prod_{i_l < i_m} V\left(p^{(i_l)}, p^{(i_m)}\right) \right\} \right) , \quad (4)$$

$$f(x;P) = 1 + \sum_{\substack{2 \leq n \leq N \\ n \text{ even}}} \left( \sum_{1 \leq i_1 < \cdots < i_n \leq N} \left\{ \prod_{j=1}^{n} \varphi\left(x; p^{(i_j)}\right) \prod_{i_l < i_m} V\left(p^{(i_l)}, p^{(i_m)}\right) \right\} \right) , \quad (5)$$

$$\varphi\left(x; p^{(i)}\right) = e^{p_\mu^{(i)} x^\mu + \delta_i} , \quad (6)$$

$$p_\mu^{(i)} p^{(i)\,\mu} = -1 , \quad (7)$$

and

$$V(p, p') = \frac{(p - p')_\mu (p - p')^\mu}{(p + p')_\mu (p + p')^\mu} = \frac{1 + p_\mu p'^\mu}{1 - p_\mu p'^\mu} . \quad (8)$$

In Eq. (6), $\delta_i$ is a constant arbitrary phase. The solitons show up in the current density:

$$J_\mu = \partial_\mu u . \quad (9)$$

**1.2.2 Properties of (1+1)-dimensional solutions**

Most of the properties of soliton solutions of SG1are shared by the solutions of SG2 and SG3. When a difference between SG1 and SG2 or SG3 exists, it is pointed out explicitly.

1) The solution, $u(x;P)$, is a function only of Lorentz scalars. As a result, it is invariant under Lorentz transformations that are applied simultaneously to the coordinate vector, $x$, and the momentum vectors. This was first pointed out in the case of the single-solution solution in Ref. [5].

2) Depending on the free phases in Eq. (6), a solution with $N \geq 2$ solitons will exhibit in the (1+$n$)-dimensional Minkowski space anywhere from one junction, when all free phases are sufficiently close to zero, and up to $N(N-1)/2$ junctions when the free phases are sufficiently large. Sufficiently far from all junctions, the current density splits up into a sum of single solitons, each associated with one of the momenta, $p^{(i)}$, used in Eqs. (2)-(8).

3) In an $N$-soliton solution, all pairs of momentum vectors obey $p^{(i)} \neq \pm p^{(j)}$ for $i \neq j$. If equality holds, then the $N$-soliton solution degenerates to a solution with a smaller number of solitons.

4) Owing to Eq. (7), the velocity of each individual soliton (be it a single-soliton solution, or one soliton in a multi-soliton solution, away from soliton junctions) is lower than the speed of light ($c = 1$). This is a consequence of the fact that a single soliton solution depends on one exponential:

$$u(x;p) = 4\tan^{-1}\left[e^{p_0 t - p_1 x + \delta}\right] . \tag{10}$$

Consequently, the velocity, $v$, of propagation of a single soliton is lower than the speed of light:

$$|v| = |p_0/p_1| = \sqrt{p_1^2 - 1}/|p_1| \leq 1 . \tag{11}$$

Hence, the Lorentz boost with a velocity equal to $v$ transforms an individual soliton to a rest frame. The solution and the associated momentum vector are transformed as follows:

$$\{p_0, p_1\} \to \{0, \pm 1\} \quad \Rightarrow \quad u(x;p) \to 4\tan^{-1}\left[e^{\pm x_1 + \delta}\right] . \tag{12}$$

Thus, in its rest fame, an individual soliton is static (stationary and time-independent). However, a transformation of any multi-soliton solution of SG1 to a rest frame does not exist, because each soliton propagates along the *x*-axis at a different velocity.

The situation is different in the case of the slower-than-light multi-soliton solutions of SG2 and SG3. Each of these solutions propagates rigidly at one velocity. Hence, it can be Lorentz transformed to a rest frame, in which it is static.

5) A Hirota-type solution of SG1 propagates at a velocity that is lower than $c$, and is stable [8]. A single-wave solution, not a Hirota soliton, which propagates at velocities that exceed $c$ also exists. This solution is unstable [8]. The present paper deals only with Hirota-type solutions in higher space dimensions. Their stability is still an open problem, as there is no extension of the Sturm-Liouville theory beyond one dimension.

**1.3 Extension of Eq. (1) to higher space dimensions**

The following Sections, present the construction of *N*-soliton solutions of Eq. (1) in (1+*n*) dimensions, $n = 2,3$, through the Hirota algorithm for all $N \geq 1$. It will be shown that:

1) In solutions of SG2 with $N \geq 3$ solitons, $(N-2)$ of the momentum vectors in Eqs. (2)-(8) are linear combinations of just two of them [33]. Each multi-soliton solution of SG2 propagates rigidly at a constant velocity. Finally, SG2 solutions are divided into two unconnected subsets. In one subset, the solutions propagate at velocities that are greater than or equal to $c$. In the other subset, they propagate at velocities that are lower than $c$. Slower-than-light solutions can be Lorentz-transformed to a rest frame, in which they are static (stationary and time-independent).

2) The multi-soliton solutions of SG3 are divided into four subsets. Two subsets are the space-rotated solutions in (1+2) dimensions. The solutions in the third and fourth subsets are genuinely three-dimensional structures – branes. Solutions in the third subset propagate rigidly at the speed of light. The fourth subset contains "hybrid" solutions, in which some clusters of solitons may propagate rigidly at velocities that are lower than $c$, but there are clusters that propagate at velocities that are greater than, or equal to $c$. Different clusters may have different velocities. One soliton may participate in more than one cluster.

3) It is impossible to reach the speed of light as a limit of slower-than-light solutions.

**2. Relativistic analysis of space-like momentum-vectors**

Owing to the space-like nature of the momentum vectors, from which Hirota-type soliton solutions are constructed, each individual soliton propagates at a velocity that is lower than $c$. However, the velocity of a cluster of solitons may be lower than, equal to, or exceed $c$. This counter-intuitive phenomenon is a direct consequence of the tachyonic nature of the momentum vectors. It does not exist exist in the case of time-like vectors, discussed in Section 6.

To classify an $N$-soliton solution, one needs to find whether clusters of two or more solitons propagate rigidly at one velocity, and what that velocity is. In particular, one needs to know if it is possible to transform a cluster of solitons, or a whole multi-soliton solution, to a rest frame, in which it is static: stationary and time independent. Such a cluster must be moving rigidly as a whole at a velocity that is lower than $c$, so that it can be Lorentz-transformed to its rest frame.

Concurrently, the momentum vectors associated with all solitons in the cluster ought to be transformed to a purely space-like form in the Lorentz frame, in which the solution is static:

$$p^{(i)} = \{p_0^{(i)}, \vec{p}^{(i)}\} \to \{0, \vec{q}^{(i)}\} \quad (\vec{q}^{(i)} \cdot \vec{q}^{(i)} = 1) \ . \tag{13}$$

In Eq. (13), $\vec{q}^{(i)}$ is a unit vector in the $n$-dimensional Euclidean space.

A Lorentz transformation in (1+3) dimensions with a boost velocity $\vec{v}$ is represented by:

$$L = \begin{pmatrix} \gamma & -\gamma\beta_x & -\gamma\beta_y & -\gamma\beta_z \\ -\gamma\beta_x & 1+(\gamma-1)\frac{\beta_x^2}{\beta^2} & (\gamma-1)\frac{\beta_x\beta_y}{\beta^2} & (\gamma-1)\frac{\beta_x\beta_z}{\beta^2} \\ -\gamma\beta_y & (\gamma-1)\frac{\beta_x\beta_y}{\beta^2} & 1+(\gamma-1)\frac{\beta_y^2}{\beta^2} & (\gamma-1)\frac{\beta_y\beta_z}{\beta^2} \\ -\gamma\beta_z & (\gamma-1)\frac{\beta_x\beta_z}{\beta^2} & (\gamma-1)\frac{\beta_y\beta_z}{\beta^2} & 1+(\gamma-1)\frac{\beta_z^2}{\beta^2} \end{pmatrix} . \tag{14}$$

In Eq. (14),

$$\vec{\beta} = \{\beta_x, \beta_y, \beta_z\} = \{v_x, v_y, v_z\}/c \ , \quad \gamma = 1/\sqrt{1-\vec{\beta}^2} \ . \tag{15}$$

In (1+2) dimensions, the matrix is reduced to its top leftmost 3×3 minor, and in (1+1) dimensions – to the 2×2 minor.

A single vector that obeys Eq. (7) can always be transformed to the form given in Eq. (13). In (1+3) dimensions, there will be a two-parameter family of transformations that will achieve this, as only one of the components of $\vec{\beta}$ can be determined. In (1+2) dimensions, this will be a one parameter family, and in (1+1) dimensions-a unique transformation.

Next, consider a pair of vectors, $p^{(1)} \neq \pm p^{(2)}$. In (1+1) dimensions, no transformation can transform both vectors to the form given in Eq. (13). There is only one free parameter, $\beta_x$, and there are two quantities, $p_0^{(i)}$, $i =1,2$ that have to be transformed to zero. There is no need to discuss the (1+3)-dimensional case, because two vectors define a plane. Hence, one can first rotate the two vectors so that their *z*-components vanish, and reduce the system to (1+2)-dimensions:

$$p^{(i)} = \{p_0^{(i)}, p_x^{(i)}, p_y^{(i)}\} \quad (i=1,2) \ . \tag{16}$$

The boost parameters, $\beta_x$ and $\beta_y$ of Eqs. (14) and (15), required for the transformed vectors to have vanishing time components, as in Eq. (13) are:

$$\beta_x = \frac{p_0^{(1)} p_y^{(2)} - p_0^{(2)} p_y^{(1)}}{p_x^{(1)} q_y^{(2)} - p_x^{(2)} p_y^{(1)}} \quad , \quad \beta_y = \frac{p_0^{(2)} p_x^{(1)} - p_0^{(1)} p_y^{(2)}}{p_x^{(1)} p_y^{(2)} - p_x^{(2)} p_y^{(1)}} \ . \tag{17}$$

For the transformation to be feasible, its velocity must be lower than *c*. Hence, the magnitude of the vector $\vec{\beta}$ must be smaller than 1. Using Eq. (17), one obtains the constraint:

$$1 - \beta_x^2 - \beta_y^2 = \frac{1 - (p^{(1)} \cdot p^{(2)})^2}{(p_x^{(1)} p_y^{(2)} - p_x^{(2)} p_y^{(1)})^2} > 0 \ . \tag{18}$$

Thus, for a pair of vectors that obey Eq. (7) to be transformable to the form given in Eq. (13), its scalar product in Minkowski space must obey

$$\left| p^{(1)} \cdot p^{(2)} \right| < 1 \ . \tag{19}$$

If the inequality is inverted,

$$\left| p^{(1)} \cdot p^{(2)} \right| \geq 1 \ , \tag{20}$$

then there is no velocity lower than *c* that can yield the desired transformation.

In the case of vectors that obey Eq. (19) (corresponding to a pair of solitons that move together at a velocity that is lower than *c*), the limit of equality,

$$p^{(1)} \cdot p^{(2)} \to \pm 1 \ , \tag{21}$$

is reached with

$$p^{(2)} \to \mp p^{(1)} \ . \tag{22}$$

This can be shown by first transforming $p^{(1)}$ and $p^{(2)}$ to the form given in Eq. (13). Eq. (21) then becomes:

$$p^{(1)} \cdot p^{(2)} = -\vec{q}^{(1)} \cdot \vec{q}^{(2)} = -\cos(\vec{q}^{(1)}, \vec{q}^{(2)}) \to \pm 1 \ . \tag{23}$$

$\cos(\vec{q}^{(1)}, \vec{q}^{(2)})$ is the cosine of the angle between the two unit vectors. The limit is obtained by:

$$\vec{q}^{(2)} \to \mp \vec{q}^{(1)} \ , \tag{24}$$

from which Eq. (22) follows in any moving frame through a Lorentz transformation.

In contrast, when Eq. (20) holds, the previous argument does not apply; one cannot transform the two vectors to the form of Eq. (13) simultaneously. The limit of Eq. (21) then consists of a continuum of vectors in (1+2) or (1+3) dimensions. A simple way to see this, is to Lorentz transform one of the vectors, say $p^{(1)}$, as in Eq. (13) (this *is* possible), with $p^{(2)}$ transformed as follows:

$$p^{(1)} \to \{0, \vec{q}^{(1)}\} \qquad (\vec{q}^{(1)} \cdot \vec{q}^{(1)} = 1) \ , \quad p^{(2)} \to \{q_0^{(2)}, \vec{q}^{(2)}\} \ . \tag{25}$$

(Note that now $\vec{q}^{(2)}$ is *not* a unit vector!) Eq. (21) then becomes:

$$p^{(1)} \cdot p^{(2)} = |\vec{q}^{(2)}| \cos(\vec{q}^{(1)}, \vec{q}^{(2)}) = \pm 1 \ . \tag{26}$$

Eqs. (26) and (7) yield:

$$|\vec{q}^{(2)}| = \frac{1}{|\cos(\vec{q}^{(1)}, \vec{q}^{(2)})|} \ , \quad q_0^{(2)} = \pm \tan(\vec{q}^{(1)}, \vec{q}^{(2)}) \ . \tag{27}$$

Thus, there is a continuum of pairs of vectors for which the limit of Eq. (21) can be achieved.

In (1+3) dimensions, there is another possibility, of simultaneously Lorentz-transforming three space-like vectors, which obey Eq. (7), to the form given by Eq. (13). Applying the Lorentz transformation of Eq. (14) to the three vectors:

$$p^{(i)} = \{p_0^{(i)}, p_x^{(i)}, p_y^{(i)}, p_z^{(i)}\} \ , \quad (i = 1, 2, 3) \ , \tag{28}$$

and requiring that the transformed vectors be of the form given in Eq. (13), one finds that the parameters of the transformation have to be:

$$\beta_x = \frac{\Delta_x}{\Delta_0} \quad , \quad \beta_y = \frac{\Delta_y}{\Delta_0} \quad , \quad \beta_z = \frac{\Delta_z}{\Delta_0} \quad . \tag{29}$$

In Eq. (29),

$$\Delta_x = \begin{vmatrix} p_0^{(1)} & p_y^{(1)} & p_z^{(1)} \\ p_0^{(2)} & p_y^{(2)} & p_z^{(2)} \\ p_0^{(3)} & p_y^{(3)} & p_z^{(3)} \end{vmatrix} \quad , \quad \Delta_y = \begin{vmatrix} p_0^{(1)} & p_z^{(1)} & p_x^{(1)} \\ p_0^{(2)} & p_z^{(2)} & p_x^{(2)} \\ p_0^{(3)} & p_z^{(3)} & p_x^{(3)} \end{vmatrix} \quad , \quad \Delta_z = \begin{vmatrix} p_0^{(1)} & p_x^{(1)} & p_y^{(1)} \\ p_0^{(2)} & p_x^{(2)} & p_y^{(2)} \\ p_0^{(3)} & p_x^{(3)} & p_y^{(3)} \end{vmatrix}$$

$$\Delta_0 = \begin{vmatrix} p_x^{(1)} & p_y^{(1)} & p_z^{(1)} \\ p_x^{(2)} & p_y^{(2)} & p_z^{(2)} \\ p_x^{(3)} & p_y^{(3)} & p_z^{(3)} \end{vmatrix} \tag{30}$$

When $\Delta_0 \neq 0$, for the transformation to be feasible, the magnitude of $\vec{\beta}$ must be smaller than 1. If $\Delta_0 = 0$, there is no solution for $\vec{\beta}$ unless $\Delta_x$, $\Delta_y$, and $\Delta_z$ vanish as well. This point will be of relevance in the classification of multi-soliton solutions in (1+3) dimensions.

**3. Soliton solutions of in (1+2) dimensions**

As will be shown in the following, in the case of $N \geq 3$ solitons, only two of the momentum vectors are independent. The remaining vectors are given as linear combinations of the two "basis" vectors, $p^{(1)}$ and $p^{(2)}$. If this pair of vectors obeys Eq. (19), then the coefficients $V(p^{(i)},p^{(j)})$ in Eqs. (4), (5) and (8) are all positive. Hence, $f(x;P)$ of Eq.( 5) does not vanish anywhere. The resulting multi-soliton solution, $u(x;P)$, varies (possibly, several times) over the range $[0, 2\pi]$. Concurrently, the solution propagates at a velocity that is lower than $c = 1$.

If $p^{(1)}$ and $p^{(2)}$ obey Eq. (20), then $f(x;P)$ of Eq.( 5) vanishes on some manifold (a line in (1+1) dimensions, a plane in (1+2) dimensions). The resulting multi-soliton solution, $u(x;P)$, varies over the range [0, 2 $N\pi$], where $N$ is the number of solitons. Concurrently, the solution propagates at a velocity that is greater than, or equal to $c$.

### 3.1 Single- and two-soliton solutions of SG2

The solution procedure of SG2 is rather cumbersome, and best implemented with the aid of symbolic manipulation software. (MATHEMATICA has been used in the case of this study). Hence, only some highlights are presented. One begins by substituting Eqs. (2)–(5) in

$$Q = \left(\partial_\mu \partial^\mu u + \sin u\right)\left(f(x;P)^2 + g(x;P)^2\right)^2 . \tag{31}$$

The single-and two-soliton solutions are readily constructed. Eqs. (6)-(8) ensure that $Q$ vanishes for these two solutions. The single-soliton solution propagates at a velocity that is lower than $c$, whereas the two-soliton solution may propagate at a velocity that is either lower than $c$ or exceeds $c$, depending in whether Eq. (19) or Eq. (20) holds, respectively, for the two momentum vectors, from which the solution is constructed.

### 3.2 Three-soliton solution of SG2

Repeating the procedure delineated above in the case of three-soliton solutions of SG2, after implementing Eqs. (6)-(8), the quantity $Q$ of Eq. (31) remains proportional to $(\Delta_z)^2$, with $\Delta_z$ defined in Eq. (30). For a three-soliton solution to exist, $\Delta_z$ must, therefore, vanish. This requirement was first observed by Hirota [25]. It means that a three-soliton solution exists only if one of the three momentum vectors, say $p^{(3)}$, is a linear combination of the other two vectors [33].

### 3.3 $N > 3$ soliton solutions of SG2

The proof that, in solutions with $N > 3$ momentum vectors, $(N-2)$ of the vectors must be linear combinations of just two of them is cumbersome but straightforward. To show how the proof goes, consider the case of $N = 4$. One repeats the construction procedure delineated above through Eqs. (2)-(8). Among the remaining monomials in $Q$ of Eq. (31), one considers the col-lection of monomials that do not $\varphi(x; p^{(i)}) = e^{p^{(i)} \cdot x}$ for some value of $1 \leq i \leq 4$. This collection is just $Q$ in the three-soliton case (i.e., as if the soliton associated with $p^{(i)}$ did not exist). For this part to vanish, the remaining three vectors, $p^{(j)}$, $1 \leq j \neq i \leq 4$, must be linearly dependent. This must hold for any $1 \leq i \leq 4$. Hence, of the four vectors, two must be linear combinations of the other two. The proof for any $N > 3$ is by induction. Of the $N$ momentum vectors, $(N-2)$ must be linear combinations of two vectors [33]:

$$p^{(i)} = \mu_i p^{(1)} + \nu_i p^{(2)} \quad (3 \leq i \leq N) \ . \tag{32}$$

### 3.4 Properties of (1+2)-dimensional solutions

As a result of Eq. (32), the velocity of a solution with $N \geq 2$ solitons is determined by the velocity required for the transformation of the two "basis" vectors (the choice of which is arbitrary), $p^{(1)}$ and $p^{(2)}$, from which all other momentum vectors are constructed, to the form given in Eq. (13).

If these two vectors obey Eq. (19), then a Lorentz transformation with a velocity, given in Eq. (17), and lower than $c$, transforms the two vectors to the form given in Eq. (13). Hence, the solution propagates as a whole at that velocity, and is transformed by the transformation to a rest frame, in which it is static: stationary and time-independent. If the two basis vectors obey Eq. (20), then the "velocity" of Eq. (17) exceeds $c$, corresponding to a solution that propagates as a whole at a velocity, $v \geq c$. Thus, the (1+2)-dimensional solutions are divided into two subsets: solutions that propagate as a whole at $v < c$ and at $v \geq c$.

The inclusion of the case of solutions whose velocity is $v = c$ in the second subset is not arbitrary. If the two basis vectors, $p^{(1)}$ and $p^{(2)}$, obey Eq. (19), then the limit of equality, Eq. (21), is achievable only through Eq. (22). However, in this limit, a slower-than-light solution degenerates into a slower-than-light solution with a smaller number of solitons. If the (+) sign holds in Eq. (22) then the coefficient $V(p^{(1)}, p^{(2)})$ of Eq. (8) vanishes, causing the solution to degenerate into a slow-er-than-light solution with (N−1) solitons. If the (−) sign holds in Eq. (22), then $V(p^{(1)},p^{(2)})$ becomes infinite, causing the solution to degenerate into a slower-than-light solution with (N−2) solitons. *In summary, the Hirota algorithm generates soliton solutions, for which the speed of light cannot be achieved from within the subset of slower-than-light solutions.*

The situation is different in the case of solutions constructed with basis vectors that obey Eq. (20). The solitons pairs then propagate rigidly at velocities that are $\geq c$. The discussion that leads to Eqs. (25)-(27) shows that the limit $v = c$ is included in this subset of solutions.

## 4. Soliton solutions in (1+3) dimensions
### 4.1 Single- and two-soliton solutions of SG3

The single- and two-solitons solutions of SG3 are readily generated by the Hirota algorithm. As in the case of the solutions of SG2, the single-soliton solution propagates at a velocity that is lower than $c$, whereas the velocity of the two-soliton solution is lower than, equal to, or higher than $c$, depending on whether the associated momentum vectors obey, respectively, Eq. (19) or Eq. (20). It is nothing but a spatially rotated (1+2)-dimensional solution.

### 4.2 $N \geq 3$ soliton solutions of SG3

Let us begin with the three-soliton solution. After having implemented Eqs. (2)-(8), one finds that for the solution to exist (namely, for $Q$ of Eq. (31) to vanish), the four determinants defined in Eq. (30) must obey the constraint presented in Ref. [26]:

$$0 = (\Delta_0)^2 - \left((\Delta_x)^2 + (\Delta_y)^2 + (\Delta_z)^2\right) =$$
$$1 - 2(p^{(1)} \cdot p^{(2)})(p^{(1)} \cdot p^{(3)})(p^{(2)} \cdot p^{(3)}) - (p^{(1)} \cdot p^{(2)})^2 - (p^{(1)} \cdot p^{(3)})^2 - (p^{(2)} \cdot p^{(3)})^2$$
. (33)

The first observation is that Eq. (33) contains the (1+2)-dimensional solutions as a special case [26]. When all vectors are (1+2)-dimensional, namely, lacking $z$-components, it yields the vanishing of $\Delta_z$ – the condition for the existence of (1+2)-dimensional solutions [25].

The second observation is that Eq. (33) limits the choice of the three momentum vectors. Viewing it as an equation for $\left(p^{(2)} \cdot p^{(3)}\right)$, the solution for the latter is:

$$\left(p^{(2)} \cdot p^{(3)}\right) = -\left(p^{(1)} \cdot p^{(2)}\right)\left(p^{(1)} \cdot p^{(3)}\right) \pm \sqrt{\left(1-\left(p^{(1)} \cdot p^{(2)}\right)^2\right)\left(1-\left(p^{(1)} \cdot p^{(3)}\right)^2\right)} \ . \tag{34}$$

As all the scalar products are real numbers, one can only have the two following possibilities. One possibility is that the three scalar products obey *simultaneously* either Eq. (19), or Eq. (20). (The proof for the existence of solutions presented in [26] applies to the case of Eq. (19).) Another possibility is that at least one of the three scalar products obeys

$$\left|\left(p^{(i)} \cdot p^{(j)}\right)\right| = 1 \ . \tag{35}$$

Despite these limitations, Eq. (33) allows for a family of three-soliton solutions that is much richer than the three-soliton solution of SG2. There are two ways to satisfy Eq. (33).

1) $\Delta_0 = 0$: Solution reduces to SG2 solution
In this situation, each of the three determinants on the r.h.s. of Eq. (33) must vanish. As a result, the three momentum vectors must be linearly dependent. One of them, say, $p^{(3)}$, must obey Eq. (32). Then, the three-soliton solution is a mere rotation into three space dimensions of a three-soliton solution of SG2. Hence again, it propagates rigidly in a plane at a velocity that is lower than, equal to, or higher than *c*, depending on whether the "basis" vectors, $p^{(1)}$ and $p^{(2)}$, obey, respectively, Eq. (19) or Eq. (20).

2) $\Delta_0 \neq 0$: Solution propagates rigidly at speed of light
The geometrical interpretation provided for this case in Ref. [26] is that Eq. (33) is the condition for the vanishing of the area of the hyper-surface spanned in Minkowski space by the three (1+3)-dimensional vectors . This situation has, in addition, a simple physical meaning. Exploiting Eqs. (29) and (30), Eq. (33) can be re-written as:

$$\beta_x^2 + \beta_y^2 + \beta_z^2 = 1 \ . \tag{36}$$

Namely, the Lorentz boost-velocity required so as to transform the three-soliton solution to a static one (the three associated momentum vectors to be transformed to the form given by Eq. (13)) is equal to speed of light. Thus, the constraint presented in Ref. [26] allows only for a non-planar three-soliton solution that propagates rigidly at the speed of light. This happens *despite* the fact that each individual soliton propagates at a velocity that is lower than $c$. The meaning of this last statement is that, if one chooses a frame of reference, in which one of the solitons is static (this is always possible), then one, or both of the other two solitons will be receding away from it at a speed that equals $c$. Such are the peculiarities of tachyonic momentum vectors…

In summary, the three-soliton solution of SG3 is either a space rotated three-soliton solution of SG2, in which case, it propagates rigidly in a plane at a velocity that is either lower than, equal to, or higher than $c$, or a genuine (1+3)-dimensional solution, which propagates rigidly at $v = c$.

### 4.2.2 Solutions with $N \geq 4$ solitons

The extension to more solitons follows the steps presented in the case of SG2. Ref. [26] provides the results for $N = 4$, and the proof for any number of solitons is, again, by induction. The remaining task is the classification of the solutions. There are four subsets of solution types.

1) $\underline{\Delta_0 = 0}$ for all momentum triplets: Solutions reduce to the two subsets of SG2 solutions

In this situation, each of the three determinants on the r.h.s. of Eq. (33) must vanish for all momentum triplets. As a result, only two of the momentum vectors are linearly independent. The remaining ($N-2$) vectors must obey Eq. (32). Then, the $N$-soliton solution is a mere rotation into three space dimensions of an $N$-soliton solution of SG2. Hence again, the solution propagates rigidly in a plane at a velocity that is lower than, equal to, or higher than $c$, depending on whether the "basis" vectors, $p^{(1)}$ and $p^{(2)}$, obey, respectively, Eq. (19) or Eq. (20).

2) $\underline{\Delta_0 \neq 0}$ for all triplets, only one linearly independent triplet: Solution propagates rigidly at $c$

Solutions that belong to this subset are three-dimensional structures – branes. The ($N-3$) momentum vectors are linear combinations of three linearly independent vectors:

$$p^{(i)} = \mu_i \, p^{(1)} + \nu_i \, p^{(2)} + \sigma_i \, p^{(3)} \quad (4 \leq i \leq N) \ . \tag{37}$$

It suffices for Eq. (33) to hold for the basis vectors. It then holds for any triplet of vectors. As a result, the whole set of $N \geq 4$ solitons propagates rigidly at the speed of light

3) $\Delta_0 \neq 0$: Hybrid solutions

When none of the conditions enumerated above, except for Eq. (33), are obeyed, the solutions are, again, three-dimensional structures – branes. They may contain mixtures of soliton clusters that propagate at velocities, which are lower than $c$, but must contain clusters whose velocity is higher than or equal to $c$. Different clusters may have different velocities. Any soliton may participate in more than one cluster. If, as a consequence of Eq. (33) every pair of momentum vectors obeys Eq. (19), then each soliton pair propagates at a speed that is lower than $c$, whereas larger clusters of solitons propagate at velocities that are either lower than, equal to, or higher than $c$. If, on the other hand, all momentum pairs obey Eq. (20), then all soliton pairs propagate at speeds that exceed $c$, and so do all larger clusters of solitons.

Example – Hybrid solution  To demonstrate the peculiarities of tachyonic momentum vectors, consider a four-soliton solution with the following momentum vectors:

$$\begin{aligned} p^{(i)} &= \{0, \cos\varphi^{(i)}, \sin\varphi^{(i)}, 0\} \ , \quad 1 \leq i \leq 3 \\ p^{(4)} &= \{p_0^{(4)}, \cos\varphi^{(4)}, \sin\varphi^{(4)}, \pm p_0^{(4)}\} \ , \quad p_0^{(4)} > 0 \end{aligned} \tag{38}$$

All triplets of vectors obey Eq. (33). Hence, this is a valid four-soliton solution in (1+3) dimensions, constructed via Eqs. (2)-(8). However, conditions 1) and 2) above are not satisfied.

The solitons constructed from $p^{(1)}$, $p^{(2)}$ and $p^{(3)}$, are static – in a rest frame in the $x$-$y$ plane. As a result, at $t = 0$, Eqs. (2)-(8) generate a four-soliton structure, three of which, (generated from $p^{(1)}$, $p^{(2)}$ and $p^{(3)}$) lie in the $x$-$y$ plane, whereas the soliton generated from $p^{(4)}$ protrudes outside this plane. Thus, the solution is a brane.

For $t \neq 0$, the solution varies in time thanks to the fact that $p_0^{(4)} > 0$. (The other three vectors have vanishing time components, hence, do not contribute to the time dependence of the solution.) A surprise is discovered in the limits $t \to \pm\infty$. For $t \to -\infty$, the solution tends to

$$4\tan^{-1}\left[\frac{e^{\xi^{(1)}} + e^{\xi^{(2)}} + e^{\xi^{(3)}} + e^{\xi^{(1)}+\xi^{(2)}+\xi^{(3)}} V(p^{(1)},p^{(2)})V(p^{(1)},p^{(3)})V(p^{(2)},p^{(3)})}{1 + e^{\xi^{(1)}+\xi^{(2)}} V(p^{(1)},p^{(2)}) + e^{\xi^{(1)}+\xi^{(3)}} V(p^{(1)},p^{(3)}) + e^{\xi^{(2)}+\xi^{(3)}} V(p^{(2)},p^{(3)})}\right], \tag{39}$$

where $\xi^{(i)}$ are the exponents in Eq. (6):

$$\xi^{(i)} = p_\mu^{(i)} x^\mu + \delta_i \quad, \quad i = 1,2,3 \ . \tag{40}$$

As $p^{(1)}$, $p^{(2)}$ and $p^{(3)}$ have vanishing time components, the result in Eq. (39) is independent of time, and represents a static three-soliton solution that lies in the $x$-$y$ plane.

The $t \to +\infty$ limit of the solution is:

$$2\pi -$$

$$4\tan^{-1}\left[\frac{\begin{pmatrix} e^{\xi^{(1)}} V(p^{(1)},p^{(4)}) + e^{\xi^{(2)}} V(p^{(2)},p^{(4)}) + e^{\xi^{(3)}} V(p^{(3)},p^{(4)}) + \\ e^{\xi^{(1)}+\xi^{(2)}+\xi^{(3)}} V(p^{(1)},p^{(2)})V(p^{(1)},p^{(3)})V(p^{(1)},p^{(4)})V(p^{(2)},p^{(3)})V(p^{(2)},p^{(4)})V(p^{(3)},p^{(4)}) \end{pmatrix}}{\begin{pmatrix} 1 + e^{\xi^{(1)}+\xi^{(2)}} V(p^{(1)},p^{(2)})V(p^{(1)},p^{(4)})V(p^{(2)},p^{(4)}) \\ + e^{\xi^{(1)}+\xi^{(3)}} V(p^{(1)},p^{(3)})V(p^{(1)},p^{(4)})V(p^{(3)},p^{(4)}) \\ + e^{\xi^{(2)}+\xi^{(3)}} V(p^{(2)},p^{(3)})V(p^{(2)},p^{(4)})V(p^{(3)},p^{(4)}) \end{pmatrix}}\right]$$

$$.\tag{41}$$

This is also a static solution, with the same three solitons. Relative to the $t \to -\infty$ limit, they are shifted by finite shifts in the $x$-$y$ plane. The shifts are consequences if the numerical coefficients that multiply each of the exponential functions. In addition, the sign of each soliton is flipped.

However, if one applies a Lorentz transformation along the $z$-axis, in the limit of the boost velocity approaching $c$ (corresponding to a reference frame that moves along the $z$-axis at a velocity equal to $\pm c$ ($z = \pm t$, the sign corresponding to $\pm p_0^{(4)}$ in Eq. (38)), the solution tends to a static four-soliton structure; the four solitons are seen as propagating rigidly at the speed of light.

## 5. Invariance property of slower-than-light solutions in (1+3) dimensions

The slower-than-light (1+3)-dimensional $N$-soliton solutions of SG3 deserve additional attention. Consider, first, a static solution. The space components of all the momentum vectors ($\vec{q}^{(i)}$ of Eq. (13)) lie in a plane in the three-dimensional space. Denote the unit vector normal to this plane by $\vec{n}$. In a moving frame, $\vec{q}^{(i)}$ are transformed to the (1+3) dimensional space-like momentum vectors $p^{(i)}$, and $\vec{n}$ is transformed into a space-like vector, $n$, obeying

$$n_\mu p^{(i)\mu} = 0 \quad , \quad p^{(i)}_\mu p^{(i)\mu} = n_\mu n^\mu = -1 \quad . \tag{42}$$

As the $x$-dependence of the solution appears only through the Lorentz invariant scalar product in Eq. (6), the soliton solutions in (1+3) dimensions are invariant under the transformation:

$$u(x;Q) = u(x + \alpha(x)n;Q) \quad , \tag{43}$$

for any scalar function $\alpha(x)$. In particular, Eq. (43) implies that the current density obeys:

$$n^\mu J_\mu = n^\mu \partial_\mu u = 0 \quad . \tag{44}$$

## 6. Extension to case of time-like momentum vectors

Consider the following modification of Eq. (1):

$$\partial_\mu \partial^\mu u - \sin u = 0, \quad \mu = 0,1,..,n, \quad n = 1,2,3 \quad . \tag{45}$$

A trivial way to obtain Eq. (45) is to replace $u$ by $(u \pm \pi)$ in Eq. (1). One constructs the solutions in the manner described in Section 1.2.1, and then adds $\pm \pi$ to the result. However, unlike the solutions of Eq. (1), the resulting solutions of Eq. (45) then do not vanish at infinity in some direction in the (1+$n$)-dimensional space. The following discussion addresses the non-trivial case, in which soliton solutions of Eq. (45) obey vanishing boundary conditions in some direction at infinity. The application of the Hirota algorithm generates $N$-soliton solutions of Eq. (45) for all $N \geq 1$, for $n = 1, 2, 3$. The only changes in the algorithm are that Eqs. (7) and (8) are replaced by:

$$p^{(i)}_\mu p^{(i)\mu} = +1 \quad , \tag{46}$$

and

$$V(p, p') = \frac{1 - p_\mu p'^\mu}{1 + p_\mu p'^\mu} \quad . \tag{47}$$

The constraints for the existence of $N \geq 3$ soliton solutions in (1+2) dimensions, (Eq. (32)), and in (1+3) dimensions (Eq. (33)), are arrived at also in the present case.

Owing to Eq. (46), in all space dimensions, an individual soliton (be it the single-soliton solution, or one soliton in a multi-soliton solution) propagates at a velocity, $v \geq c$. The (1+1)-dimensional solutions are readily constructed. Interesting effects of the time-like nature of the momentum vectors on multi-soliton solutions occur in higher space dimensions. The effects are consequences of the fact that there is no Lorentz transformation that can simultaneously transform two, or more, *different* time-like vectors to a rest frame (certainly, no transformation to Eq. (13) exists!).

(1+2) dimensions
The two-soliton solution is comprised of two solitons, each propagating at a different velocity, $v \geq c$. In solutions with $N \geq 3$ solitons, $\Delta_z$, defined in Eq. (3), must vanish for all triplets of momentum vectors. Hence, only two vectors are linearly independent, and all remaining vectors must obey Eq. (32). Again, each soliton propagates at a different velocity, $v \geq c$.

(1+3) dimensions
In the two-soliton solution, each soliton propagates at a different velocity, $v \geq c$. Hence, is a rotated (1+2)-dimensional solution. All (1+3)-dimensional solutions with $N \geq 3$ solitons are also mere space rotations of (1+2)-dimensional solutions. The reason is that all momentum vectors must obey Eq. (32). Namely, there are only two linearly independent vectors.

This conclusion is a direct consequence of the time-like nature of the momentum vectors, combined with the fact that all momentum triplets must obey Eq. (33), with $\Delta_0$, $\Delta_x$, $\Delta_y$ and $\Delta_z$ defined in Eq. (30). To see this, consider the triplet of vectors in Eqs. (30) and (33). One can always Lorentz transform one of them, say $p^{(3)}$, to its rest frame:

$$p^{(3)} \to \{1,0,0,0\} \ . \tag{48}$$

As a result, $\Delta_0$ of Eq. (30) vanishes in the rest frame. Obeying Eq. (33) then requires that each of the three determinants, $\Delta_x$, $\Delta_y$ and $\Delta_z$ must vanish. This, in turn, requires that the space parts of $p^{(1)}$ and $p^{(2)}$ must be proportional to one another:

$$p^{(1)} = \{p_0^{(1)}, \vec{p}^{(1)}\} \ , \quad p^{(2)} = \{p_0^{(2)}, \alpha\, \vec{p}^{(1)}\} \ . \tag{49}$$

Eqs. (48) and (49) yield a linear relation amongst the (1+3)-dimensional vectors:

$$p^{(3)} = \frac{\left(\alpha\, p^{(1)} - p^{(2)}\right)}{\alpha\, p_0^{(1)} - p_0^{(2)}} \ . \tag{50}$$

This linear relation is preserved when the three vectors are transformed back to their forms prior to the transformation that yields Eq. (48). As this conclusion applies to all momentum triplets, only two of them are linearly independent.

Acknowledgments The author wishes to thank G. Bel, G.I. Burde, I Rubinstein, and B. Zaltzman for many constructive comments.